\documentclass[showpacs,aps,floatfix,superscriptaddress]{revtex4}
\usepackage{amsmath,amssymb,eucal,graphicx,subfigure}
\usepackage{color}
\begin{document}

\def\a{\alpha}
\def\b{\beta}
\def\g{\gamma}
\def\gin{\g_{\rm in}}
\def\gout{\g_{\rm out}}
\def\Pin{P_{\rm in}}
\def\Pout{P_{\rm out}}
\def\k{\kappa}

\def\av#1{\langle#1\rangle}
\def\be{\begin{equation}}
\def\ee{\end{equation}}

\title{Growing networks with superjoiners}

\author{Ameerah Jabr-Hamdan}
\email{jabrhaal@clarkson.edu}
\affiliation{Department of Physics, Clarkson University, Potsdam, NY 13699-5820}
\author{Jie Sun}
\email{sunj@clarkson.edu}
\affiliation{Department of Mathematics \& Computer Science, Clarkson University, Potsdam, NY 13699-5815}
\author{Daniel ben-Avraham}
\email{benavraham@clarkson.edu}
\affiliation{Department of Physics, Clarkson University, Potsdam, NY 13699-5820} 
\affiliation{Department of Mathematics \& Computer Science, Clarkson University, Potsdam, NY 13699-5815}

\begin{abstract}
We study the Krapivsky-Redner (KR) network growth model but where new nodes can connect to any number of existing nodes, $m$, picked from a power-law distribution $p(m)\sim m^{-\a}$. Each of the $m$ new connections is still carried out as in the  KR model with probability redirection $r$ (corresponding to degree exponent $\gamma_{\rm KR}=1+1/r$ in the original KR model).  The possibility to connect to any number of nodes resembles a more realistic type of  growth in several settings, such as social networks, routers networks, and networks of citations.  Here we focus on the  in-, out-, and total-degree distributions and on the potential tension between the degree exponent $\a$, characterizing new connections (outgoing links), and the degree exponent $\g_{\rm KR}(r)$ dictated by the redirection mechanism.
\end{abstract}

\pacs{%
89.75.Hc,  
02.50.-r   
}
\maketitle

\section{Introduction}
Complex networks have garnered much recent attention for their intrinsic mathematical interest and their many applications to everyday life~\cite{albert02,newman03,barrat08,dorogovtsev08}.  A particular area of research is the study of simple  growth models that capture the networks' more salient features~\cite{watts98,barabasi99,dorogovtsev00,jin01,growthbook}.  A well known example of this is
the Krapivsky-Redner  growth redirection model (KR)~\cite{kr}, where each new node is connected to a randomly selected node, with probability $1-r$, or else the connection is ``redirected" to the ancestor of that node, with probability $r$ (the {\it ancestor} of $x$ is the node that $x$ connects to upon joining the network).   
The redirection events select preferentially for existing nodes of higher degree (the rich-get-richer effect), giving rise to a {\em scale-free} degree distribution with degree exponent $\g_{\rm KR}=1+1/r$ --- arguably, the most central characteristic of real-life complex networks.  Moreover, the KR model achieves this with a simple, {decentralized} algorithm (only local knowledge about the target node is required at each step) that in all likelihood captures an essential ingredient in the growth of real complex networks: befriending a friend of a friend, in social networks, or finding a new reference through an already cited paper, in a network of citations, are obvious analogs of the KR redirection mechanism.  On the other hand, it is unrealistic to expect, in real-life examples, that each joining agent makes only {\em one} connection.

A result of the restriction to one connection is that the KR model yields loopless (acyclic) graphs, or {\em trees}, which simplifies their theoretical analysis, but is also in disparity with real-life complex networks that tend to have abundant loops and a high degree of clustering~\cite{ebel02,hidalgo08,bullmore09,kaluza10,loopy}.  In~\cite{rb} we explored the possibility of connecting each new joining node to $m$ of the existing network nodes, with probability $p_m$, where each of the $m$ connections follows the original KR redirection recipe.  If the support of $m$ is finite, $m=1,2,\dots,m_{max}$, then the resulting networks are still scale-free,
with the same degree exponent $\g_{\rm KR}=1+1/r$, but they now contain loops and their degree distributions for small $k$ may be conveniently manipulated through an appropriate choice of the $p_m$'s~\cite{rb}.  Our interest here is in the case that the $p_m$ do {\em not} have finite support, but $m_{max}$ grows along with the size of the network, and we focus on power-law distributions of the form $p_m\sim m^{-\a}$, such that the probability for {\it superjoiners} --- new nodes that connect to most of the nodes already in the network --- is substantial. 
This power-law choice is motivated by several practical situations.  For example, in social networks it is well known that the distribution of the number of acquaintances of a person is governed by a power-law, and it is then plausible that the number of  friends $m$ a person would make upon joining a new social network would be power-law distributed as well;
when a company introduces a new router to the Internet, we may expect that it would be connected to $m$ routers, with a power-law distribution, since larger companies can afford more connections in proportion to their resources, and companies' sizes follow a power-law distribution, etc.  

There is a potential tension between the exponent $\a$ governing the distribution of {\em outgoing} links of each new node (we regard the links formed from each new node as {\em directed} from the node to their target), and the exponent $\gin=1+1/r$ that one expects for the {\em incoming} links, that form under the KR rich-get-richer mechanism with redirection probability~$r$.  How does this tension play out and which of the two effects dominates the {\em total} degree distribution, where all the links of a node are counted, regardless of their directionality?  Below we shall see  that rich behavior results from this interesting tug of war.  
We show that the degree distribution of outgoing links is $\Pout(l)\sim l^{-\gout}$, $\g_{out}=\a$, whereas that of incoming links is $\Pin(k)\sim k^{-\gin}$, where $\gin=1+1/r$ for $\a>2$, and  $\gin=0$ (a flat distribution) for $\a<2$.
 The distribution of the total degree (in- {\em and} out-links) is $P(q)\sim q^{-\g}$, where $\g=\min\{\gin,\gout\}$ for $\a>1$, and $\g=-(1-\a)$ for $\a<1$. In the latter case, $\g<0$, with nodes of larger degree becoming more abundant, in complete reversal from the usual state of affairs for everyday complex nets.  

\begin{figure}[h]
\includegraphics[width=0.45\textwidth]{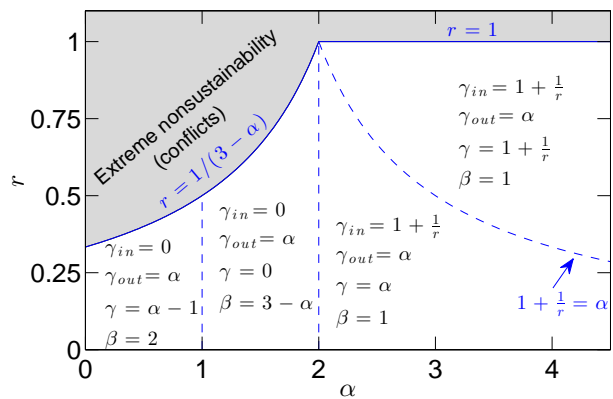}
\caption{(Color online) ``Phase diagram" for the superjoiners model:  The model exhibits different
asymptotic behavior and divides into four regions in the $(\a,r)$ space, with regards to the exponents characterizing the asymptotic power-law behavior of in-, out-, and total-degree distribution ($\gin$, $\gout$, and $\g$, respectively).  For $\a>2$, the networks are sustainable, $M\sim N^{\beta}$, $\beta=1$, and $\g=\min\{\gin,\gout\}$, yielding two distinct ``phases" separated by the curve $\gout=\gin$, or $\a=1+1/r$.  For $\a<2$, the networks are unsustainable, with $\beta>1$, and the region divides into two distinct ``phases" separated by the line $\a=1$.  Additionally, the superjoiners model is ill-defined in the region marked as ``extreme non-sustainability," for $r>1/(3-\a)$, as conflicts arise in the implementation of the superjoiners growth model.  The various results summarized in this diagram are derived throughout the paper.}
\label{phase_diagram.fig}
\end{figure}

One result of the fat-tailed distribution for the number of new links, $m$, is that the total number of links in the network, $M(t)$, may grow faster than the number of nodes, $N(t)$.  We find that $M(t)\sim N(t)^\b$, with $\b=1$ for $\a>2$,  $\b=3-\a$ ($>1$) for $1<\a<2$, and $\beta=2$ for $\a<1$.  The case in which $\b>1$ and $M(t)$ outpaces the growth of $N(t)$ is known as the {\em non-sustainable} regime.  In fact, due to the fast growth of $M$ conflicts might arise on introducing a new node, when some of its random connections call to be directed to a common target.  For $\a<2$ and $r>1/(3-\a)$ we find a regime of {\em extreme non-sustainability}, where the number of conflicts becomes a {\em finite} fraction of the total number of links and our model becomes ill-defined.
The rich behavior of the superjoiners model is summarized in Fig.~\ref{phase_diagram.fig}.  Throughout the remainder of this paper we present the theoretical and numerical considerations that led us to these conclusions.
 
 An infinite support for $p_m$ was already  briefly considered in~\cite{rb}, for the ``self-consistent" case that it is dictated by the network's own degree distribution, $p_m=P(m)$. In that case the total degree distribution is scale-free, with $\g=1+1/r$.  Krapivsky and Redner~\cite{copying} have made a detailed study of network growth by ``copying," where new nodes connect to a randomly selected node and to all its ancestors, so that there too $m$ may grow without bound.  Power-law distribution for new connections (with unbounded $m$) were studied by Tessone et al.,~\cite{tessone},  in the context of dependency networks in Open Source Software projects, where they observed $\a$ values in the range of $2.2$--$3.5$. Although their theoretical analysis focuses around strictly linear growth kernels (yielding BA-like models limited to $\g=3$), many of our findings here are closely related to theirs. 
 
\section{Model description}

Our network model is constructed by adding one node at a time.  A newly added node, $y$, is connected to $m$ of the existing nodes in the network, $x_1,x_2,\dots,x_m$.  The connections are {\it directed}, pointing from $y$ into each of the $m$ target nodes.  The $m$ target nodes are called the {\it ancestors} of $y$.  Each of the $m$ target nodes is selected by the Krapivsky-Redner recipe: A node $x$ is selected at random and it is identified as $x_1$ with probability $1-r$.  Otherwise, with probability $r$, $x_1$ is a randomly selected ancestor of $x$.  In either case, a directed link is created from $y$ to $x_1$.  This process is repeated $m$ times, until all target nodes are determined (and $y$ has $m$ out-links and $m$ ancestors).  For now, we assume that {\it conflicts}---when the same target node is selected for more than one of the $m$ connections---do not arise, or at least are rare enough to be neglected.  Conflicts are addressed in Section~\ref{conflicts.sec}.

The number of nodes in the system grows by $1$ with the addition of each new node, so that the total number of nodes 
after step $t$ is 
\be
N(t)=N(0)+t\,,
\ee
where $N(0)$ is the initial number of nodes at the starting configuration, at time $t=0$.  The range of $m$, therefore, may increase linearly with time.  In this work, we are interested in what happens when $m$ is selected from a power-law distribution, such as
\be
\label{pm.eq}
p_m(t)=C(t)m^{-\a}\,,\qquad m=1,2,\dots,N(t)-1\,,
\ee
where both the normalization factor, 
\be
\label{ct.eq}
C(t)=1\Big/\sum_{m=1}^{N(t)-1}m^{-\a}\,,
\ee 
and the range of $m$ adapt to the growing network~\cite{remark1}.  
Although $m$ can grow without bound,  it is still bounded at any {\it finite} time.  This allows us, in principle, to consider any value for the exponent $\a$, including cases in which $C(t)^{-1}\to\infty$, as $t\to\infty$.  We nevertheless limit our study to $\a>0$, since unsustainability becomes extreme as $\a$ decreases, and it is hard to think of  practical examples for $\a<0$.
The normalization factor $C(t)$ converges to a finite value for $\a>1$, but not for $\a<1$, resulting in the asymptotic behavior
\be
\label{pm_asym.eq}
p_m(t)\sim m^{-\a}\times\begin{cases}
1 & \a>1,\\
t^{-(1-\a)} & \a<1.
\end{cases}
\ee

The starting configuration, at time $t=0$, has no significant effect on the network at large times. For concreteness, however, our simulations (and numerical integrations of the theory's equations) start with two nodes, $v_A$ and $v_B$, connected to one another via directed links: $v_A\to v_B$ and $v_B\to v_A$.  From now on, we specialize our discussion to this case, where $N(0)=2$, $M(0)=2$, and $N(t)=t+2$.

\section{theory and results}
Let $N_{kl}(t)$ denote the number of nodes with $k$ links in and $l$ links out, at time $t$.  The total number of nodes with in-degree $k$ (regardless of the out degree) is $F_k(t)=\sum_lN_{kl}(t)$, and the number of nodes with out-degree $l$ (regardless of the in-degree) is $G_l(t)=\sum_kN_{kl}(t)$.  The number of nodes with {\em total} degree $q$ is
\be
\label{Hq.eq}
H_q(t)=\sum_{k,l\atop k+l=q}N_{kl}(t)\,.
\ee
The total number of nodes in the network at time $t$ is  
\be
\sum_kF_k=\sum_lG_l=\sum_qH_q=N(t)\,.
\ee
The {\it average number of outgoing links} added to the net at time $t$ is
\be
\label{kappa.eq}
\k(t)=\sum_{m=1}^{t+1}mp_m(t)\,,
\ee
or, in the asymptotic limit of $t\gg1$,
\be
\label{kappa_asym.eq}
\k(t)\sim\begin{cases}
\k_{\infty}, & \a>2,\\
t^{2-\a}, & 1<\a<2,\\
t, & \a<1,
\end{cases}
\ee
where we have made use of (\ref{pm_asym.eq}), and $\k_{\infty}$ is a constant (dependent on $\a$).

The {\it expected total number of outgoing links} at time $t$ is then
\be
\label{M.eq}
M(t)=M(0)+\sum_{t'=1}^t\k(t')\,,
\ee 
where $M(0)$ is the number of outgoing links in the initial configuration. Note that the total number of incoming links and outgoing links is equal at all times (and equal to the total number of links, since each link is simultaneously an in-, and out-link).  Using Eq.~(\ref{kappa_asym.eq}) and the fact that $N(t)\sim t$ for all $\a$, we get
\be
\label{M_asym.eq}
M(t)\sim N(t)^{\b}\,;\qquad \b=\begin{cases}
1, & \a>2,\\
3-\a, & 1<\a<2,\\
2, &\a<1.
\end{cases}\ee
Thus, for $\a<2$, the growth of links outpaces that of the nodes, and our model then yields {\em unsustainable} networks, with $\b>1$. 

$N_{kl}$ satisfies the rate equation:
\be
\label{Nkl.eq}
N_{kl}(t+1)-N_{kl}(t)=\k(t)\frac{1-r}{N(t)}[N_{k-1,l}(t)-N_{kl}(t)]+\k(t)\frac{r}{M(t)}[(k-1)N_{k-1,l}(t)-kN_{kl}(t)]
+p_l(t)\delta_{k,0}\,.
\ee
The first term on the right-hand side denotes the changes to $N_{kl}$ due to direct connections: a node is selected for such a process with probability $(1-r)/N(t)$, and this takes place $\k(t)$ times during each step (on average).  The second term denotes redirected connections: a node of in-degree $k$ (and out-degree $l$) is selected with probability $kN_{kl}/M(t)$.  The last term accounts for the new node added to the net: the new node has $l$ outgoing links (and zero incoming links), with probability $p_l(t)$.
Equation~(\ref{Nkl.eq}) is supplemented with the boundary condition
\be
 N_{kl}(t)=0 \qquad {\rm for\ }k<0{\rm\ or\ }l<1\,.
\ee

\subsection{Outgoing links}
\label{outgoing.sec}
\begin{figure}[h]
\includegraphics[width=0.45\textwidth]{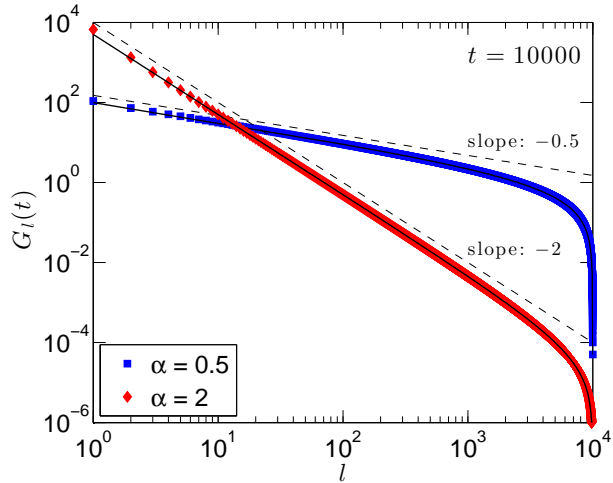}
\caption{(Color online) Out-degree $G_l(t)$ for  $\a=0.5$  (upper curve) and $\a=2$ (lower curve), for $t=10000$.  The symbols represent a direct summation of Eq.~(\ref{Gl.eq}) and the solid curves are fit from Eq.~(\ref{Gl_asymp.eq}). Slopes of $l^{-\a}$ are shown as broken lines, for comparison.}
\label{Gl.fig}
\end{figure}

Summing (\ref{Nkl.eq}) over $k$, we obtain $G_l(t+1)-G_l(t)=p_l(t)$,
or
\be
\label{Gl.eq}
G_l(t)=G_l(0)+\sum_{t'=0}^{t-1}p_l(t')=2\delta_{l,1}+\sum_{t'=l-1}^{t-1}p_l(t')\,.
\ee
[The lower limit of $t'=l-1$ in the last sum is explained by the fact that $p_l(t)=0$ for $l>t+1$.]
This can be obtained more directly upon realizing that the out-degree of each node is determined as it is added to the network, and never changes thereafter.  If $\a>1$, the normalization factor for $p_l(t)$ converges with time to a finite value, while it vanishes as $C(t')\sim t'^ {\a-1}$ for $\a<1$, leading to 
\be
\label{Gl_asymp.eq}
G_l(t)\sim l^{-\a}\times\begin{cases}
(1-l/t), & \a>1\,,\\
(1-(l/t)^{\a}), & \a<1\,.
\end{cases}
\ee
In either case, $G_l(t)\sim l^{-\a}$ for finite $l/t$, as illustrated in Fig.~\ref{Gl.fig}.  This means that $\gout=\a$
for all values of $\a$ (and $r$).

\subsection{Incoming links}
\label{incoming.sec}
Summing (\ref{Nkl.eq}) over $l$, we obtain
\be
\label{Fk.eq}
F_k(t+1)-F_k(t)=\k(t)\frac{1-r}{N(t)}[F_{k-1}(t)-F_k(t)]+\k(t)\frac{r}{M(t)}[(k-1)F_{k-1}(t)-kF_k(t)]+\delta_{k,0}\,.
\ee
The outcome now depends on the time-asymptotic behavior of $N(t)$, $\k(t)$, and $M(t)$.  For~$\a>2$, $\k(t\to\infty)=\k_{\infty}$ converges to a constant, and $M(t)\sim\k_{\infty}t$.  Using these asymptotic relations, along with $N(t)\sim t$ and the {\it ansatz} $F_k(t)\sim f_kt$ (where $f_k$ is a constant independent of time), one gets
\be
\label{fk.eq}
f_k=\k_{\infty}(1-r)[f_{k-1}-f_k]+r[(k-1)f_{k-1}-kf_k]+\delta_{k,0}\,.
\ee
This can be analyzed exactly~\cite{kr}, leading to $f_k\sim k^{-(1+1/r)}$, or $\gin=1+1/r$ for $\a>2$.

\begin{figure}[h]
\includegraphics[width=0.45\textwidth]{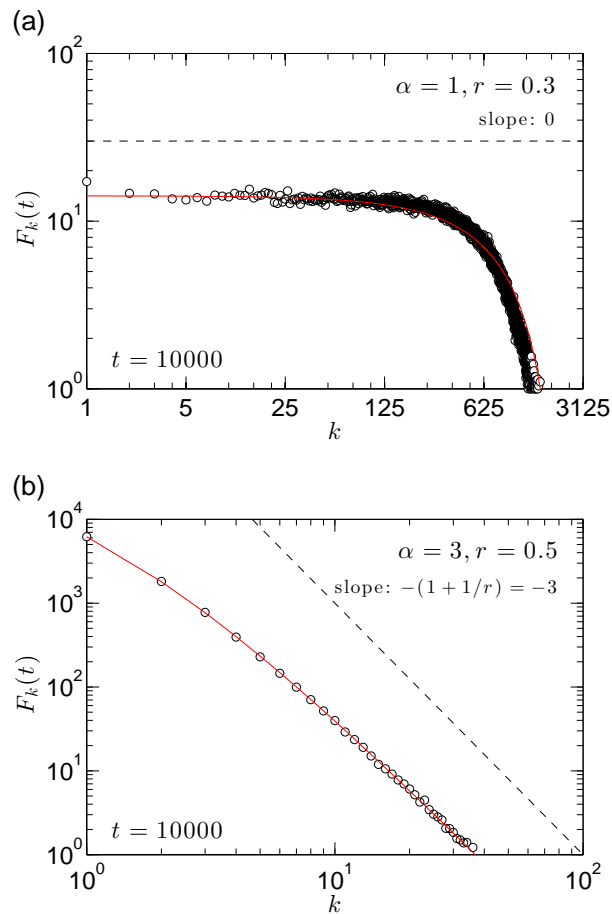}
\caption{(Color online) In-degree $F_k(t)$ for $\a=1$, $r=0.3$~(a) and $\a=3$, $r=0.5$~(b), as computed from Eq.~(\ref{Fk.eq}) (solid line) and from simulations ($\circ$) for $t=10000$, averaged over 50 runs.  The predicted asymptotic slopes, of $\gout=0$ and $\gout=1+1/r=3$, respectively, are shown for comparison (broken line).}
\label{Fk.fig}
\end{figure}

For $\a<2$, the term with $\k(t)/N(t)$ dominates Eq.~(\ref{Fk.eq}), asymptotically, and determines the outcome, which now becomes $F_k(t)\to{\rm const.}$, independent of $k$.  In other words,
$\gin=0$ for $\a<2$.  The two regimes for $\gin$ ($\a>2$ and $\a<2$) are demonstrated in Fig.~\ref{Fk.fig}.

\subsection{All links}
\label{all.sec}
To derive $H_q(t)$, we need to deal fully with Eq.~(\ref{Nkl.eq}).  This can be done exactly by exploiting the fact that the index $l$ is fixed and it effectively enters the equation only as a boundary condition.  First, set $k=0$ to obtain
\be
\label{Nk0.eq}
N_{0l}(t+1)-N_{0l}(t)\left[1-\k(t)\frac{1-r}{N(t)}\right]=p_l(t)\,,
\ee
which can be solved with the help of the integrating factor
\be
\label{A.eq}
A_0(t)=\prod_{j=1}^t\left[1-\k(j)\frac{1-r}{N(j)}\right]^{-1}\,,
\ee
to yield
\be
\label{N0lsol.eq}
N_{0l}(t)=A_0(t-1)^{-1}\sum_{i=0}^{t-1}A_0(i)p_l(i)\,.
\ee
In general, for $k\geq1$, Eq.~(\ref{Nkl.eq}) can be rewritten as
\be
\label{Nkla.eq}
N_{kl}(t+1)-N_{kl}(t)\left[1-\k(t)\left(\frac{1-r}{N(t)}+\frac{rk}{M(t)}\right)\right]=\k(t)\left[\frac{1-r}{N(t)}+
\frac{r(k-1)}{M(t)}\right]N_{k-1,l}(t)\,,
\ee
which can be solved for $N_{kl}(t)$, in terms of $N_{k-1,l}(t)$, using an integrating factor similar to $A_0(t)$,
\be
\label{Ak.eq}
A_k(t)=\prod_{j=1}^t\left[1-\k(j)\left(\frac{1-r}{N(j)}+\frac{rk}{M(j)}\right)\right]^{-1}\,,
\ee
leading to
\be
\label{Nklsol.eq}
N_{kl}(t)=A_k(t-1)^{-1}\sum_{i=0}^{t-1}A_k(i)\k(i)\left[\frac{1-r}{N(i)}+
\frac{r(k-1)}{M(i)}\right]N_{k-1,l}(i)\,.
\ee
Thus, one can obtain $N_{1l}$ from the known $N_{0l}$, then $N_{2l}$ from $N_{1l}$, etc.
This procedure, however, is cumbersome and does not allow for a simple analysis of asymptotic properties.
Instead, we observe that the weak interaction between $k$ and $l$ should lead to a near absence of correlations between the in-and out-degrees, so that 
\be
\label{correlations.eq}
N_{kl}(t)\approx F_k(t)G_l(t)/N(t)\,, 
\ee
where the denominator is dictated by normalization.
Beyond the theoretical insights gained by this simplification, the uncorrelated in- and out-degrees allow for numerical integration of much larger networks: from algorithms that grow as $N^3$, for integration of $N_{kl}$ directly [Eq.~(\ref{Nkl.eq})], to algorithms of order $N^2$ for the integration of both $G_l$ and $F_k$ [Eqs.~(\ref{Gl.eq}) and (\ref{Fk.eq}].

Assuming noncorrelation and in view of~(\ref{Hq.eq}), the probability for total degree $q$ can be written as a convolution, 
\be
\label{Hq_approx.eq}
H_q(t)\approx\sum_{k}F_k(t)G_{q-k}(t)/N(t)\,.
\ee
  If we further use the long-time asymptotic results,
$F_k\sim k^{-\gin}$ and $G_l\sim l^{-\gout}$ then, approximating the sum with an integral and analyzing the divergences at the lower and upper limits, we find
\be
\label{Hq_slopes.eq}
H_q\sim q^{-\g}\sim\begin{cases}
q^{-\min\{\gin,\gout\}}, & \gin>1,\>\gout>1,\\
q^{-\gin}, & \gin<1,\>\gout>1,\\
q^{-\gout}, & \gin>1,\>\gout<1,\\
q^{1-\gin-\gout}, & \gin<1,\>\gout<1.
\end{cases}
\ee
The first case corresponds to $\a>2$, where $\gin=1+1/r\,(>1)$ and $\gout=\a\,(>1)$. Thus, for $\a>2$ we have
$\g=\min\{1+1/r,\a\}$ and the region includes two phases (with distinct $\g$) separated by the curve $\a=1+1/r$.  On that curve,
$\g=\gin=\gout$.  The second case corresponds to the region $1<\a<2$, where $\gin=0\,(<1)$ and $\gout=\a\,(>1)$. Hence, for $1<\a<2$ we have $\g=0$ and the line $\a=2$ demarcates a sharp transition for both the values of $\gin$ (from $1+1/r$ for $\a>2$, to $0$ for $\a<2$) and of $\g$ (from $\a$ to $0$).  The third case does not occur in our model. Finally, the fourth case corresponds to $\a<1$, where $\gin=0\,(<1)$ and $\gout=\a\,(<1)$.  In this region, we get the strange result of a negative $\g$ exponent: $\g=-(1-\a)$.  An accessible
summary of the different regimes for the  predicted values of  $\gin$, $\gout$, and $\g$, as a function of $\a$ and $r$, is presented in the ``phase diagram" of Fig.~\ref{phase_diagram.fig}.

\begin{figure}[h]
\includegraphics[width=0.45\textwidth]{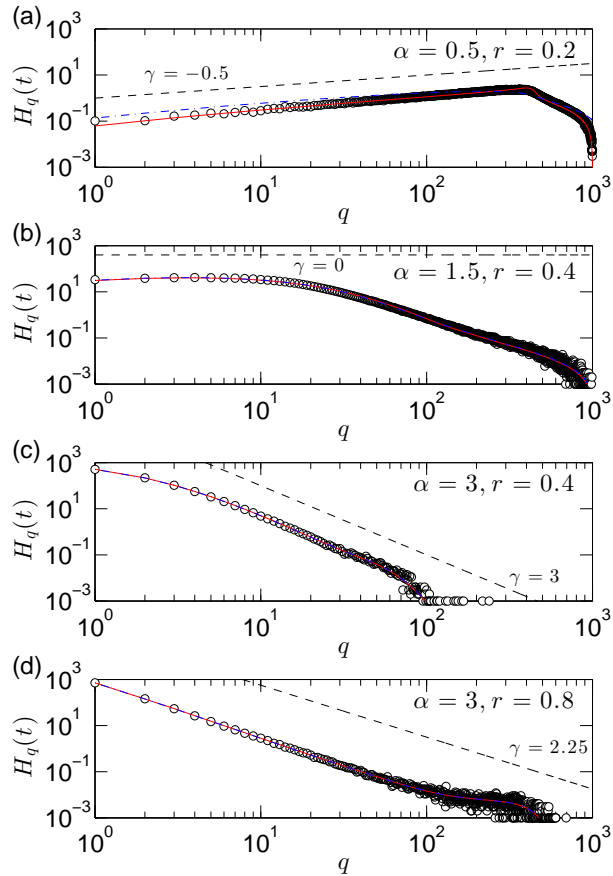}
\caption{(Color online) Total degree distribution $H_q(t)$ in small networks of $t=1000$ for (a)~$\a=0.5$, $r=0.2$; (b)~$\a=1.5$, $r=0.4$; (c)~$\a=3$, $r=0.4$; and (d)~$\a=3$, $r=0.8$. The solid curves represent the direct analytical approach~(i), while dashed-dotted curves are computed from approach (ii).  Simulation results, averaged over 1000 runs, are denoted by open circles ($\circ$). For comparison, the theoretical long-time asymptotic prediction of the exponent $\g$ is indicated by dashed lines.}
\label{Hq.fig}
\end{figure}

We have two theoretical approaches for the computation of $H_q(t)$: (i)~Compute $N_{kl}(t)$ {\em directly} from integration of Eq.~(\ref{Nkl.eq}) and then use Eq.~(\ref{Hq.eq}), and (ii)~Integrate Eqs.~(\ref{Gl.eq}) and (\ref{Fk.eq}) numerically to obtain $G_l(t)$ and $F_k(t)$, respectively; then assuming {\em noncorrelation}, use Eq.~(\ref{Hq_approx.eq}).  We next wish to compare these theoretical approaches to one another as well as to simulation results.  In Fig.~\ref{Hq.fig} we show results for representative $(\a,r)$ pairs in each of the  four regions of the phase diagram (Fig.~\ref{phase_diagram.fig}).  The direct theoretical approach~(i) is shown as a solid curve, and the noncorrelation approach~(ii) is shown as dash-dotted curve, while the simulation results are denoted by symbols.  Because (i) can be carried out only for relatively small nets, we limit ourselves to $t=1000$.  It is encouraging that the long-time asymptotic prediction for $\g$ (shown as dashed lines) provides a reasonable description for even such small nets. In fact, we find that the range where the asymptotic behavior applies increases with the size of the net.
The good agreement between  (i) and  (ii) (the curves are practically indistinguishable, other than in the first panel) supports the noncorrelation approximation, which we used for the derivation of $\gamma$.  The results also demonstrate that either theoretical approach, (i) or (ii), provides a very apt description of the  transient behavior observed in small nets.

\subsection{Conflicts and extreme unsustainability}

\label{conflicts.sec}

We now address the possibility of {\it conflicts}, when the same target node is selected more than once  and multiple links between pairs of nodes may result (for $m>1$).  Conflicts may be dealt with in several practical ways: (a)~When a conflict arises, that is, when a target is selected twice, repeat the random selection process until a new (non-conflicting) target is found. (b)~When a conflict arises, do not connect.  This means that the actual number of connections for the new node might be smaller than $m$.  (c)~When a conflict arises, make the connection regardless of the conflict, thus allowing for multiple directed links between pairs of vertices. 
\begin{figure}[htbp]
\includegraphics[width=0.45\textwidth]{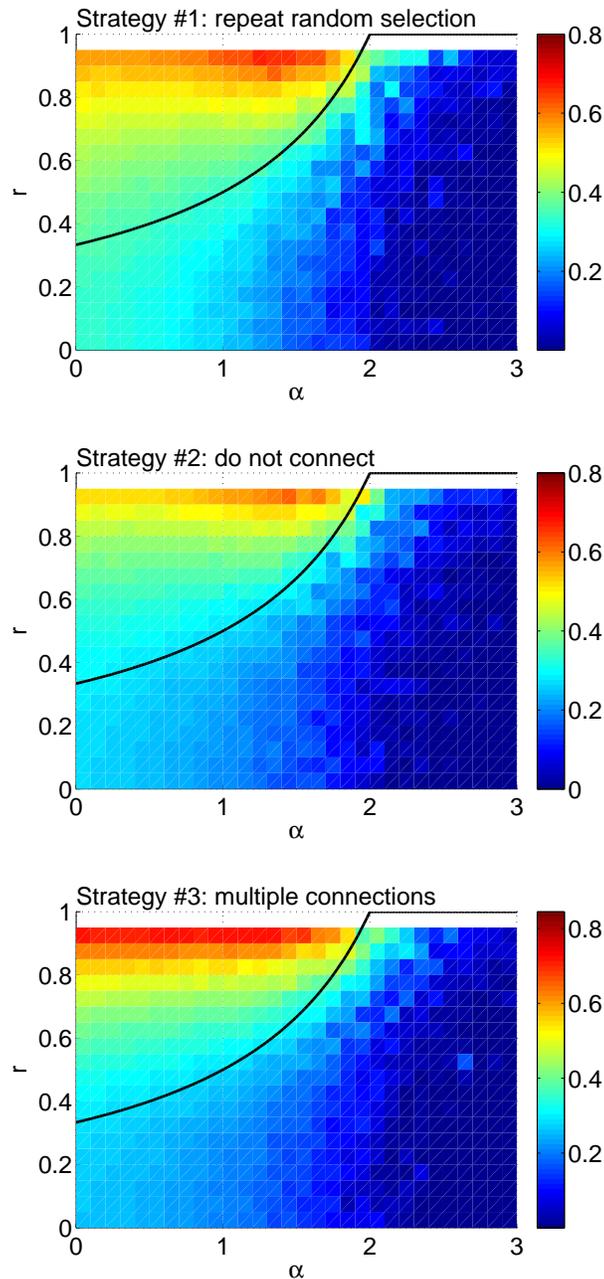}
\caption{(Color online) Conflicts. The fraction of conflicts that arise in the construction of networks up to time $t=5000$ for each of the three strategies described in the text is indicated in different shadings, according to the scheme of the bars at the right.  The theoretical curve $r=1/(3-\a)$ reasonably separates between regions of many conflicts (extreme unsustainability) and almost no conflicts in all three cases. }
\label{conflicts.fig}
\end{figure}

In the original KR model, where $m$ is always 1, conflicts never arise.  If $\kappa(t)$ tends to a {\it finite} number as $t\to\infty$, as is the case for $\a>2$, conflicts are rare in our model and they have a negligible influence on the outcome: it does not matter then which strategy one adopts to deal with conflicts.  
For $\alpha<2$, however,
conflicts are no longer rare and deviations between the results from the various conflict strategies might be expected.  

Generally, we expect conflicts to be relevant if they occur as a {\em finite fraction} of all attempts to connect new links.  If the ratio of conflicts to the total number of attempts tends to zero, in the thermodynamic limit of $t\to\infty$, then they can be neglected.  Indeed, our principal equation~(\ref{Nkl.eq}) for the evolution of $N_{kl}(t)$ does not take into account the possibility of conflicts, so it is valid only in the latter case.  Its domain of applicability is suggested by the iterative solution of Eq.~(\ref{Nklsol.eq}): A necessary requirement for the solution to make sense is that  $\left[1-\k(t)\left(\frac{1-r}{N(t)}+\frac{rk}{M(t)}\right)\right]>0$ for all $k\leq t$.
Putting $k=t$, and using the asymptotic expressions for $N(t)$, $\k(t)$, and $M(t)$, we get
\be
1-\k(t)\left(\frac{1-r}{N(t)}+\frac{rt}{M(t)}\right)\sim
\begin{cases}
1-r, &\a>2,\\
1-(3-\a)r, &1<\a<2,\\
1-\frac{1-\a}{2-\a}(1-r)-2r, &\a<1.
\end{cases}
\ee
The first case, for $\a>2$, satisfies the requirement for all values of $r$, consistent with the expectation that conflicts are negligible for $\a>2$.  Surprisingly, the two other cases,
for $1<\a<2$ and for $\a<1$, yield the very same condition: 
\be
\label{demarcate.eq}
r<\frac{1}{3-\a}\qquad {\rm for\ }\a<2\,.
\ee
Thus the regime of $r>1/(3-\a)$, $\a<2$ is not only {\em unsustainable} ($\beta>1$), but one expects a finite fraction of conflicts there.  We term this phenomenon {\em extreme unsustainability}.  In Fig.~\ref{conflicts.fig} we plot the fraction of conflicts in simulations, using all three strategies.  The boundary suggested by 
Eq.~(\ref{demarcate.eq}) seems to capture the transition to extreme sustainability quite adequately.  

\subsection{Graph spectra of superjoiners networks}
Finally, having established the effect of superjoiners on the degree distribution, we address the question of how they affect the dynamics of processes on such networks. Toward that end, we focus on the spectral properties of the adjacency matrix $A=[A_{ij}]_{N\times N}$ and the Laplacian matrix $L=[L_{ij}]_{N\times N}$, as these play a key role in the interplay between structure and dynamics in networks in general~\cite{barrat08,dorogovtsev08,arenas08,nishikawa10,skardal14prl}. The adjacency matrix is defined as
\begin{equation}
A_{ij}=
\begin{cases}
1 & \mbox{if node $i$ connects to node $j$},\\
0 & \mbox{otherwise,}
\end{cases}
\end{equation}
and the Laplacian matrix $L$ is given by $L_{ij}=\delta_{ij}d_i-(1-\delta_{ij})A_{ij}$, where $d_i=\sum_{k}A_{ik}$ is the out-degree of node $i$.

Based on the construction of the superjoiners networks, $A$ has the lower-triangular block form
\begin{equation}\label{Adj.eq}
A=\begin{pmatrix}
\begin{array}{c|c}
  A^{(0)} & O \\ \hline
  A^{(10)} & A^{(1)} 
\end{array}
\end{pmatrix},
\end{equation}
where $A^{(0)}$ is the adjacency matrix of the initial network of $N(0)$ nodes,
$A^{(1)}$ is the adjacency matrix among the added nodes, and $A^{(10)}$ encodes the directed links between those two groups. Since no self-loops are allowed, both $A^{(0)}$ and $A^{(1)}$ have zero diagonals. Moreover, since each new node can only connect to nodes that joined the network before it, the matrix $A^{(1)}$ is in fact lower-diagonal. Consequently, the spectrum of $A$ is given by
\begin{equation}
	\Lambda(A) = \Lambda(A^{(0)})\cup\{0,0,\dots,0\},
\end{equation}
where $\Lambda(M)$ denotes the set of eigenvalues of a matrix $M$. 
The same argument applies to the Laplacian matrix $L$, which yields the spectrum
\begin{equation}
	\Lambda(L) = \Lambda(L^{(0)})\cup\{d_{N(0)+1},d_{N(0)+2},\dots,d_N\},
\end{equation}
where $L^{(0)}$ is the Laplacian matrix of the initial network.

In the examples considered in this paper, the initial network contains $N(0)=2$ interconnected nodes. It follows that $\Lambda(A^{(0)})=\{-1,1\}$ and consequently $\Lambda(A)=\{-1,0,0,\dots,0,1\}$. The eigenvalues of $A$ have been theoretically hypothesized and experimentally confirmed to be a determining factor for neuronal activity~\cite{kinouchi06,larremore11}. In fact, the dynamic range, quantifying the range of stimuli that results in distinguishable responses, was shown to be maximized when the largest eigenvalue of $A$ (in magnitude) equals 1~\cite{kinouchi06,larremore11}, as in our case.

Turning to the Laplacian matrix $L$, it follows that for the initial network, $\Lambda(L^{(0)})=\{0,2\}$ while the rest of the eigenvalues of $L$ are given by the out-degrees of the remaining nodes in the network. Since the out-degrees of added nodes are bounded by $1\leq d\leq N-1$, Eq.~(\ref{pm.eq}), the eigenvalues of $L$ can be ordered as
\begin{equation}
	0=\lambda_1(L)<1=\lambda_2(L)\leq\dots\leq\lambda_N\leq N-1.
\end{equation}	

\begin{figure}[htbp]
\includegraphics[width=0.45\textwidth]{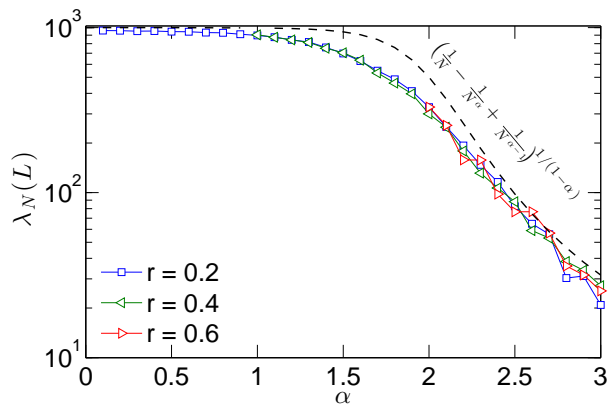}
\caption{(Color online) Largest Laplacian eigenvalue $\lambda_N(L)$ as a function of $\alpha$ in the superjoiners model averaged over $100$ network simulations per parameter combination. The dashed curve shows the analytical estimate given by Eq.~\eqref{loglambda.eq}.}
\label{spectra.fig}
\end{figure}

As an example of the use of  $\Lambda(L)$, consider the ratio $\lambda_N(L)/\lambda_2(L)$:  The synchronizability of a network of identical dynamical units is generally enhanced the smaller this ratio~\cite{pecora98}. 
For our superjoiners networks, $\lambda_2(L)=1$, as the first node added always has a single outgoing link. 
On the other hand, the value of $\lambda_N(L)$ generally decreases as a function of $\alpha$ and shows essentially no dependence on the reduction probability $r$. In fact, the dependence of $\lambda_N(L)$ on $\alpha$ can be estimated, approximately, from extreme-value statistics. A new node that joins at time $t$ can connect to up to $N(t)-1$ nodes.  If we make the simplifying assumption that each node can connect to $N$, the final (larger) network size, then we would be overestimating $\lambda_N(L)$.  With that assumption, we get 
\begin{equation}\label{loglambda.eq}
	\lambda_N(L)\approx \left(\frac{1}{N}-\frac{1}{N^\a}+\frac{1}{N^{\a-1}}\right)^{1/(1-\a)}\sim
	\begin{cases}
	N,&\a\ll1,\\
	N^{1/(\a-1)},&\a\gg1.
	\end{cases}
\end{equation}
Since the outgoing links statistics is not affected by $r$, neither is the value of $\lambda_N(L)$. In Fig.~\ref{spectra.fig} we show simulation  results for $\lambda_N(L)$ of superjoiners nets of different $\alpha$, along with the prediction from extreme statistics.  The figure confirms our analysis, and it shows that the synchronizability is better the higher the value of $\a$ is.

\section{Discussion}

Traditional network growth models allow each incoming node to connect to a capped number of existing nodes.
In this paper, we explored the possibility that the number of new connections $m$ is uncapped, but instead grows with the network size $N(t)$, and where $m$ is taken from a power-law distribution $P(m)\sim m^{-\a}$.  Keeping our network model design close to the KR growth model~\cite{kr} has allowed us to explore analytically the competition between the exponent $\a$, characterizing the degree of new nodes, and the expected degree exponent $\g_{\rm KR}=1+1/r$ that arises from the rich-get-richer bias in the original KR model.  For $\a>2$, our network model is sustainable ($M\sim N$),
the in-degree exponent is $\gin=\a$, the out-degree exponent is $\gout=1+1/r$, and the total-degree exponent $\g=\min\{\gin,\gout\}$.  For $\a<2$, the superjoiners network model is unsustainable ($M\sim N^{\beta}$, $\beta>1$) and the various exponents compete in interesting ways, as summarized in Fig.~\ref{phase_diagram.fig}.  The superjoiners network model is extremely unsustainable, in the sense that conflicts that arise when trying to connect new nodes become common-place and analytically intractable, for $\a<2$ and $r>1/(3-\a)$.

Aside from the long-time asymptotic power-law distributions of the in-, out-, and total-degree, which we were able to derive analytically, the model exhibits rich transient behavior.  Numerical integration of the master equation~(\ref{Nkl.eq}) predicts very nicely the results from simulations, but because of the three independent variables ($k$, $l$, and $t$) and computer time and memory constraints, this procedure is limited to rather small nets.  On the other hand, we have shown that the in- and out-degrees correlate only weakly, and $N_{kl}(t)$ can then
be obtained from $F_k(t)$ and $G_l(t)$, allowing for easier analysis and numerical integration of substantially larger nets.  A most important open question regarding transient behavior is deriving the specific ranges (of $k$, $l$, and $q$) where the long-time asymptotic predictions are valid. We expect that the noncorrelation phenomenon would be a great help in finding the answer.

\acknowledgements
This work is partially supported by the Simons Foundation Grant No.~318812 (J.~S.).


\begin{thebibliography}{99}
\bibitem{remark1} For simulations (and numerical integration of the equations) we actually use $p_m=C(t)'\int_{m-1/2}^{m+1/2}x^{-\a}\,dx$, $C(t)'=1/\int_{1/2}^{t+3/2}x^{-\a}\,dx$, since it makes for simpler programming code.  For large $m$, the differences
between the two distributions are small and turn out to be irrelevant to the model's characteristic behavior.
\end{thebibliography}
\end{document}